\UseRawInputEncoding
\documentclass[a4, 10 pt, conference]{ieeeconf}  
\usepackage[dvipdfmx]{graphicx}

\usepackage{here}
\usepackage{float}
\IEEEoverridecommandlockouts                              

\title{\LARGE \bf
Proposal of a dialogue system using Mecab
}

\author{Goro Miyano$^{1}$ , Kotaro Matsuo$^{2}$ , Tomohiro Kawatsu$^{2}$ , Ibuki Unno$^{2}$, Kisumi Shibata$^{2}$,Mitsuki Fumoto$^{2}$
\thanks{$^{1}$Information Science College Department of Information Security}%
\thanks{$^{2}$Information Science College Department of Practical AI Course}%
}

\begin{document}

\maketitle
\thispagestyle{empty}
\pagestyle{empty}


\begin{abstract}

In recent years, artificial intelligence (AI) has developed in a variety of
fields and is now integrated into many aspects of our daily
lives. This includes AI speakers, communication robots, and other
interactive systems. We are interested in AI speaker-like systems
in which the robot responds to the user’s needs and speech
synthesis. We participated in an interactive robot competition
to promote the development of interactive robots.

\end{abstract}

\section{Program Composition}

\subsection{Competition Overview}
This competition is the second edition of the world’s first interactive robot competition held in 2020 to compare the performance of interactive robots using androids that closely resemble humans [1],[2]. The robot acts as a counter-sales person, helping the customer decide on a tourist destination in an area where the travel agency is located. The customer chooses two of the six possible destinations before interacting with the robot, which describes the highlights of the two locations, answers questions, and helps the customer choose which of the two destinations to visit. One of the two tourist attractions chosen by the customer is randomly determined as the recommended tourist attraction. Moreover, changes in the degree of customers’ desire to visit the recommended attraction before and after the dialogue are subject to evaluation. The tasks required of the robot are as follows:
\begin{itemize}
\item Conduct a conversation that satisfies the customer.
\item Explain tourist attractions and information.
\item Provide appropriate responses to customer questions.
\item Recommend sightseeing spots in places where the customer wants to visit.
\end{itemize}

\section{SYSTEM CONFIGURATION}

Middleware provided by the convention secretariat was used for face recognition, speech recognition and synthesis, robot facial expression generation, gaze control, lip movement generation, head movement generation, and posture control. The program development environment is described below. 
Development environment,
\begin{itemize}
\item OS Windows 10
\item IDE Eclipse Visual Studio Code
\item Language Python 3.7.9 Java 1.8.0 \_206
\item Library Mecab Gensim
\item Language model jawiki.word \_vectors.300d
\end{itemize}

\begin{figure}[tbh]
\begin{center}
\includegraphics[scale=0.25]{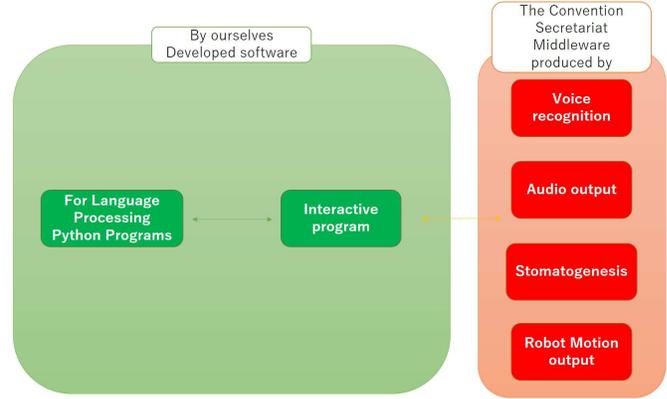}
\caption{System Configuration}
\label{fig:iruka}
\end{center}
\end{figure}

\section{PROGRAM PURPOSE}

The is program aims to familiarize the customer familiar with the robot, so that he/she will pay attention to the robot’s arguments.

\section{DIALOGUE SCENARIOS}

The dialogue scenario for this competition was set up as follows:
Greetings, Ice Breaker → Impressive Tourist Attractions → Explanation of Tourist Attraction A, Reception of Questions → Explanation of Tourist Attraction B, Reception of Questions → Presentation of Recommended Tourist Attractions

\begin{figure}[tbh]
\begin{center}
\includegraphics[scale=0.4]{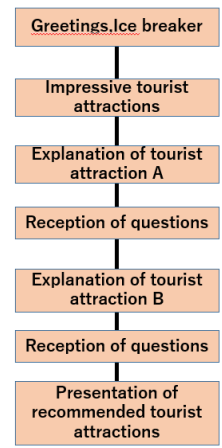}
\caption{Dialogue scenario}
\label{fig:iruka}
\end{center}
\end{figure}

\section{LANGUAGE PROCESSING TECHNIQUES}

\subsection{Linguistic processing of the most memorable tourist attractions}

For the linguistic processing of the most remarkable tourist attractions, we used the method of extracting proper nouns by morphological analysis using Mecab.

For the linguistic processing of the most remarkable tourist attractions, we used the method of extracting proper nouns by morphological analysis using Mecab.

Example (Customer) 私が最も印象に残っている観
光地は京都です (My most memorable sightseeing spot is Kyoto） → (after proper noun extraction by Mecab) 京都 (Kyoto). 
If the customer does not reply, then the value is temporarily set to “そこ” (default setting).

\begin{figure}[h]
\begin{center}
\includegraphics[scale=0.25]{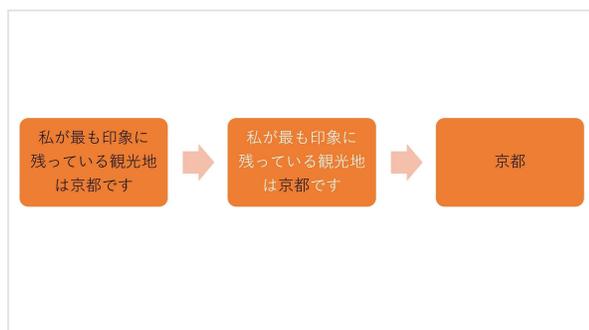}
\caption{Extraction of proper nouns}
\label{fig:iruka}
\end{center}
\end{figure}

\subsection{Linguistic processing of questions about tourist attractions}

For the linguistic processing of questions about tourist attractions, we divided the customers’ questions into seven categories according to the tourist attraction information provided to the convention office. These include price, opening hours, opening days, station, highway, car-parking, and no question. We adopted  two approaches from there. The first was to divide the customer’s question into categories by setting up, in advance, the words for which a specific category would be selected. 
For example, if the word “料金” (admission fee) is included in the question, then the fee category is selected; if the word “ 電車 ” (train) is included, then the station category is selected; if “特にありません” (nothing in particular) is included, then the no question category is selected, and so on.
The second approach was to measure the similarity between sentences using Word Rotator’s Distance (WRD)[3]. Four sentences were set for each of the seven categories. The category to which the closest sentence among the 28 sentences belonged was chosen as the category for that question.

\begin{figure}[t]
\begin{center}
\includegraphics[scale=0.3]{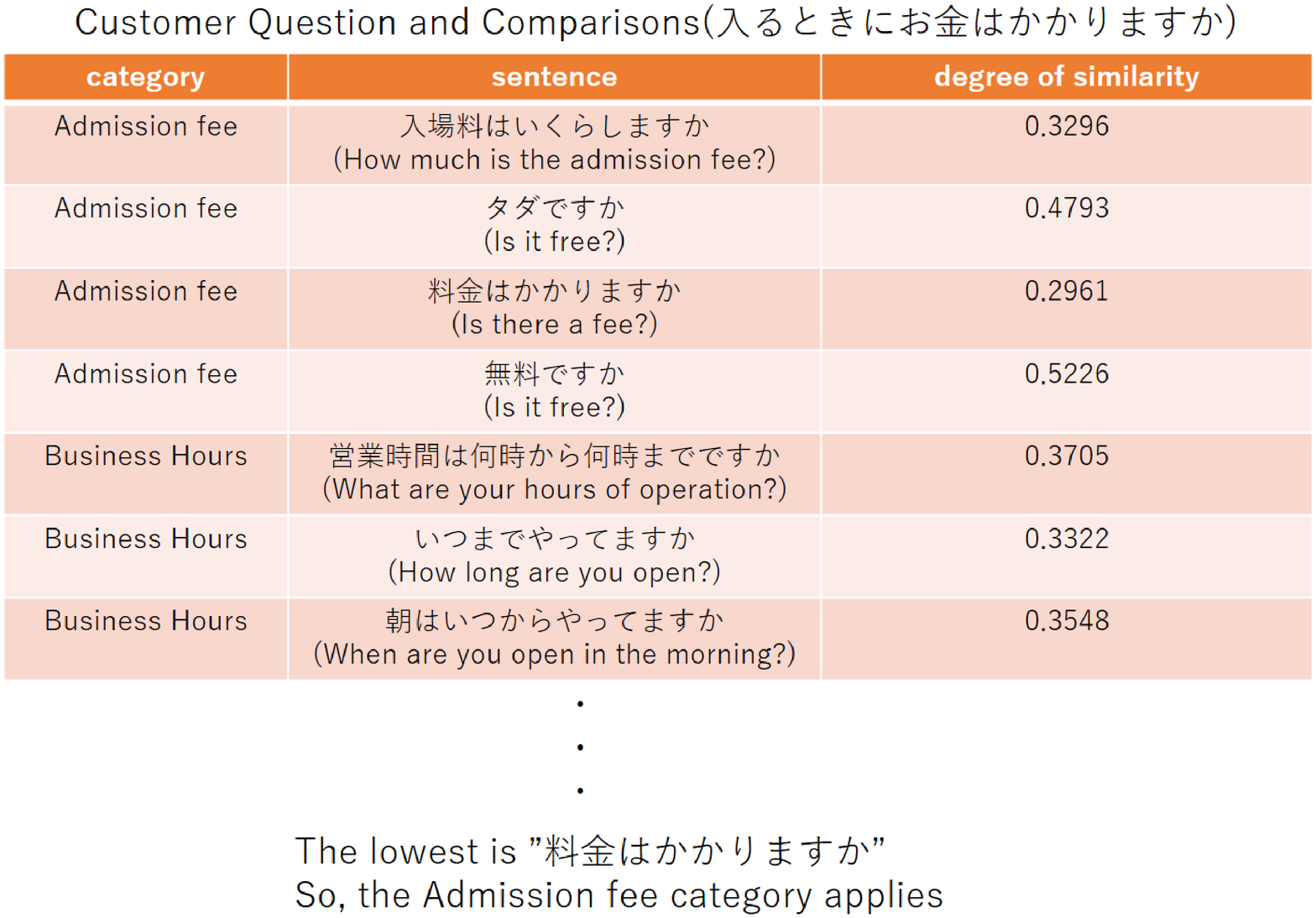}
\caption{WRD Usage Example}
\label{fig:iruka}
\end{center}
\end{figure}




\section{Robot motions}

In this system, the robot performs a “nodding” motion and turns toward the monitor where the sightseeing spot is displayed. The nodding motion is performed when the robot asks the customer for a response, and this motion is not performed when the robot is speaking.

\section{DEVICES, ETC.}

\subsection{Greetings and Ice Breaker}

Miraikan (the venue for the preliminary round) is brought up as an icebreaker as the venue is a common topic of conversation with guests.

\subsection{Familiarizing the customers}
As we could not shake the impression that robots are merely that, robots, and not like humans, we decided to make them sound familiar by daring them to speak in a “robot-character” manner. Specifically, after asking about the sightseeing spots that left an impression on the customer, we asked the customer again about the same spot in depth and then had the robot say, “I see. I updated my memory about (the sightseeing spot that left an impression on the customer). The robot is made to utter the words “I see.” 
Additionally, we made the robot say “master” once at the end when it was supposed to say “customer.”

\subsection{Nodding}
The robot does not make a nodding motion while it is in the middle of a conversation. This is to avoid a conversation conflict with the customer, and the nodding should only be done when the customer asks for an answer.

\section{Results and Discussion}

\subsection{Result}

Regarding the evaluation method, this qualifying round is evaluated based on two factors: impression evaluation items and recommended effects. The team with the highest overall evaluation are eligible to participate in the main line. The results of this team’s qualifying round are shown below.

\begin{table}[h]
\caption{Impression evaluation items}
\label{table_example}
\hspace{1cm}
\begin{tabular}{|c||c|}
   \hline
   evaluation item & Rating (average of 7-point scale)\\
   \hline \hline
   Have you been able to choose a tourist destination to visit with satisfaction? & 3.055556 \\
   Did you hear enough information about tourist attractions? & 2.722222 \\
   Were you able to interact with the robot naturally? & 2.277778 \\
   Was the robot's response appropriate? & 2.611111 \\
   Was the robot's response favorable? & 3.055556 \\
   Were you satisfied with your interaction with the robot? & 3.166667 \\
   Did you trust the robot? & 3.111111 \\
   Did you use the information obtained from the robot to help you choose a tourist destination? & 3.833333 \\
   Would you visit this travel agency again? & 3.111111 \\
   \hline
\end{tabular}
\end{table}

\subsection{Recommendation effect}

The degree to which the experienced person wants to visit the “recommended sightseeing spot” changed before and after the dialogue is evaluated numerically (-100 to 100). Notably, the recommended effectiveness score was 9.888889.

\subsection{Consideration}

The impression evaluation items all scored in the low 2s to low 3s on a seven-point scale, especially the questions (1) “Did you get enough information about the sights?” and (2) “Did you have a natural dialogue with the robot?” were notably low. We believe this is because the robot could not sufficiently answer the sightseeing spot questions asked by the customers. Regarding the recommendation effect, it can be said that it falls within the range of the top three teams to the bottom runner-up group, which does not mean that they are far away from the other teams compared to the low impression evaluation item. This is probably because the evaluation item of recommendation effectiveness is not related to the level of perfection unless the system is particularly excellent. In fact, the top three teams had higher impression evaluation items and higher recommendation scores than the other teams, but eight of the other ten teams’ recommendation scores were within the range of 5 to 15 points, and even the teams with low impression evaluation items outperformed the teams with high recommendation scores. Moreover, the results show that even teams with low impression rating items outperformed teams with high recommendation effectiveness.

\section{FUTURE ISSUES AND IMPRESSIONS}

\subsection{Issues to be addressed in the future}
As the impression evaluation items are generally low, we believe the program needs to be fundamentally revised. First, the accuracy of the WRD was not very high and often misidentified question categories, so we will consider ways to improve the accuracy or use other methods. Also, since of categories. In addition, we would like to rework the conversation scenario, as asking about “memorable sightseeing spots” felt too abrupt. 
Furthermore, regarding “robot-like” speech and behavior, we would like to hear more people’s stories to make it more familiar, while at the same time, we would like to see some parts of the speech that are more human-like, such as the strength of the speech and related robot movements. In addition, we would like to address the robot's inability to process the main question even if it picks up on the customer’s responses during the conversation, which is then reflected in the language processing section. References are important to the reader. Therefore, each citation must be complete and correct. If\\\\\\\\\\\\\\\\\\\\\\\\\ at all possible, references should be commonly available publications.


References are important to the reader; therefore, each citation must be complete and correct. If at all possible, references should be commonly available publications.

\end{document}